\documentstyle[prl,aps]{revtex}
\input epsf
\draft

\ifx\epsffile\undefined
\newlength{\epsfysize}
\def\epsffile#1#2#3#4]#5{}
\fi

\begin{document}

\newcommand{\be}{\begin{equation}}
\newcommand{\ee}{\end{equation}}
\newcommand{\bear}{\begin{eqnarray}}
\newcommand{\eear}{\end{eqnarray}}
\newcommand{\gsim}{\lower.7ex\hbox{$\;\stackrel{\textstyle>}{\sim}\;$}}
\newcommand{\lsim}{\lower.7ex\hbox{$\;\stackrel{\textstyle<}{\sim}\;$}}
\newcommand{\dr}{\mbox{\footnotesize{$\overline{\rm DR}$\ }}}
\newcommand{\GeV}{{\rm GeV}}
\newcommand{\TeV}{{\rm TeV}}
\newcommand{\met}{\not\!\!\!E_T}
\newcommand{\msusy}{M_{\rm SUSY}}
\def\npb#1 #2 #3 #4 {Nucl.~Phys. {\bf B#1}, #2 (#3)#4 }
\def\plb#1 #2 #3 #4 {Phys.~Lett. {\bf B#1}, #2 (#3)#4 }
\def\prd#1 #2 #3 #4 {Phys.~Rev.  {\bf D#1}, #2 (#3)#4 }
\def\prl#1 #2 #3 #4 {Phys.~Rev.~Lett. {\bf #1}, #2 (#3)#4 }
\def\pr#1  #2 #3 #4 {Phys.~Rept. {\bf #1}, #2 (#3)#4 }
\def\mpl#1 #2 #3 #4 {Mod.~Phys.~Lett. {\bf A#1}, #2 (#3)#4 }
\def\zpc#1 #2 #3 #4 {Z.~Phys. {\bf C#1}, #2 (#3)#4 }


\title{Supersymmetry Signatures with Tau Jets at the Tevatron}

\author{Joseph~D.~Lykken and Konstantin T. Matchev}

\address{Theoretical Physics Department, 
         Fermi National Accelerator Laboratory, 
         Batavia, IL 60510, USA}

\maketitle

\begin{abstract}
We study the supersymmetry reach of the Tevatron in channels
containing both isolated leptons and identified tau jets.
In the most challenging case, where the branching ratios of gauginos
to taus dominate, we find that searches for two leptons,
a tau jet and a large amount of missing transverse energy
have a much better reach than the classic trilepton signature.
With total integrated luminosity of ${\rm L} \gsim
4\ {\rm fb}^{-1}$, the Tevatron will start extending
the expected LEP-II reach for supersymmetry.
\end{abstract}
\pacs{PACS numbers: 12.60.Jv, 14.80.Ly, 13.35.Dx
\hfill FERMILAB-PUB-99/034-T
}


\setcounter{footnote}{0}
\setcounter{page}{1}

Searches for supersymmetry (SUSY) in Run I of the Tevatron
have been done exclusively in channels involving some
combination of leptons, jets, photons and missing transverse
energy ($\met$) \cite{Tevatron searches}. At the same time, several Run I
analyses have identified hadronic tau jets, e.g. in $W$-production
\cite{W to tau} and top decays \cite{top to tau}. 
Hadronic taus have also been used to place limits on
a charged Higgs \cite{H+ to tau} and leptoquarks \cite{Leptoquark to tau}.
As tau identification is expected to improve further in Run II,
this raises the question whether SUSY searches in channels involving
tau jets are feasible.

SUSY signatures with tau leptons are very well motivated, since
they arise in a variety of models of low-energy supersymmetry, e.g.
gravity-mediated \cite{BCDPT-PRL,BCDPT-PRD,JW} or the minimal
gauge-mediated models \cite{JW,DN,BT}. In this paper we shall
study all possible {\em experimental}
signatures with three identified objects
(leptons or tau jets) plus $\met$, and compare their reach
to the clean trilepton channel \cite{Barger,Run I 3L},
which is the classic SUSY signature at the Tevatron.
It arises in the decays of gaugino-like
chargino-neutralino pairs $\tilde\chi^\pm_1 \tilde \chi^0_2 $. 
The reach is somewhat limited by their rather small leptonic
branching fractions. In the limit of either heavy or equal in mass
squarks and sleptons, the
leptonic branching ratios of $\tilde\chi^\pm_1$ and $\tilde\chi^0_2$
are $W$-like and $Z$-like, respectively. However, both
gravity-mediated and gauge-mediated models of SUSY breaking
allow the sleptons to be much lighter than the squarks,
thus enhancing the leptonic branching fractions of the gauginos.

There are at least three generic reasons as to why one may expect light
sleptons in the spectrum. First, the slepton masses at the high-energy
scale may be rather small to begin with.
This is typical for gauge-mediated models, since the sleptons are
colorless and do not receive large soft mass contributions
proportional to the strong coupling constant $\alpha_s$. 
The minimal gravity-mediated (mSUGRA) models, on the other hand,
predict light sleptons if the universal scalar mass $M_0$ is
much smaller than the universal gaugino mass $M_{1/2}$.
Second, the renormalization group equations for the scalar soft masses
contain terms proportional to Yukawa couplings, which tend to reduce 
the corresponding mass during the evolution down to low-energy scales.
This effect is significant for third generation scalars, and
for large values of $\tan\beta$ (the ratio of the Higgs vacuum
expectation values $v_2$ and $v_1$) splits the staus
from the first two generation sleptons. And finally,
the mixing in the charged slepton mass matrix further reduces the
mass of the lightest eigenstate. The slepton mixing is enhanced at
large $\tan\beta$, since it is proportional
to $\mu m_l\tan\beta/m^2_{\tilde l}$, where 
$m_l$ ($m_{\tilde l}$) is the lepton (slepton) mass and
$\mu$ is the supersymmetric Higgs mass parameter.
Notice that this effect again
only applies to the staus, since $m_\tau\gg m_{\mu,e}$.

Due to these three effects, it may very well be
that among all scalars, only the lightest sleptons
from each generation (or just the lightest stau $\tilde\tau_1$)
are lighter than $\tilde\chi^\pm_1$ and $\tilde \chi^0_2$.
Indeed, in both gravity-mediated and gauge-mediated models
one readily finds regions of parameter space where either 
$m_{\tilde\chi^0_1}< m_{\tilde \tau_1}\sim
m_{\tilde \mu_R}< m_{\tilde\chi^+_1}\sim m_{\tilde\chi^0_2}$
(typically at small $\tan\beta$) or
$m_{\tilde\chi^0_1}< m_{\tilde \tau_1}<
m_{\tilde\chi^+_1}\sim m_{\tilde\chi^0_2} < m_{\tilde \mu_R}$
(at large $\tan\beta$). Depending on the particular model,
and the values of the parameters, the gaugino pair
decay chain may then end up overwhelmingly
in {\em any one} of the four final states: $lll$, $ll\tau$, $l\tau\tau$
or $\tau\tau\tau$.

In order to make a final decision as to which experimental signatures
are most promising, we have to factor in the tau branching ratios
to leptons and jets. About two-thirds of the subsequent tau decays
are hadronic,
so it appears advantageous to consider signatures with tau jets
in the final state as alternatives to the clean trilepton signal.
(From now on, we shall use the following terminology:
a ``lepton'' ($l$) is either a muon or an electron;
a tau is a tau-lepton, which can later decay either leptonically,
or to a hadronic tau jet, which we denote by $\tau_h$.)
The presence of taus in the underlying SUSY signal
always leads to an enhancement of the signatures with tau jets
in comparison to the clean trileptons. This disparity is most striking
for the case of $\tau\tau\tau$ decays, where
$BR(\tau\tau\tau\rightarrow ll\tau_h)/
 BR(\tau\tau\tau\rightarrow lll)\sim 5.5$.
An additional advantage of the tau jet channels is
that the leptons from tau decays are much softer than the tau jets
and as a result will have a relatively low reconstruction efficiency.

On the other hand, the tau jet
channels suffer from larger backgrounds than the clean trileptons.
The physical background (from {\em real} tau jets in the event)
is actually smaller, but a significant part of the background is due to
events containing narrow isolated QCD jets with the correct track
multiplicity, which can be misidentified as taus. 
The jetty signatures are also hurt by
the lower detector efficiency for tau jets than for leptons.
The main goal of our study, therefore, will be to see what is the
net effect of all these factors, on a channel by channel basis.

For our analysis we choose to examine one of the most challenging
scenarios for SUSY discovery at the Tevatron.
We shall assume the typical large $\tan\beta$ mass hierarchy
$m_{\tilde\chi^0_1}< m_{\tilde \tau_1}<
m_{\tilde\chi^+_1}< m_{\tilde \mu_R}$. One then finds that
$BR(\tilde\chi^+_1\tilde\chi^0_2\rightarrow \tau\tau\tau+X)\simeq 100\%$
below $\tilde\chi^\pm_1\rightarrow W^\pm \tilde\chi^0_1$
and   $\tilde\chi^0_2\rightarrow Z\tilde\chi^0_1$ thresholds.
In order to shy away from specific model dependence, we shall
conservatively ignore all SUSY production channels other than
$\tilde\chi^\pm_1\tilde \chi^0_2 $ pair production.
The $p_T$ spectrum of the taus resulting from the chargino and
neutralino decays depends on the mass differences
$m_{\tilde\chi^+_1}-m_{\tilde\tau_1}$ and
$m_{\tilde\tau_1}-m_{\tilde\chi^0_1}$.
The larger they are, the harder the spectrum, and the better the
detector efficiency. However, as the mass difference gets large,
the $\tilde\chi^+_1$ and $\tilde\chi^0_2$ masses themselves
become large too, so the production cross-section is severely suppressed.
Therefore, at the Tevatron we can only explore regions with favorable
mass ratios and at the same time small enough gaugino masses.
This suggests a choice of SUSY mass ratios: for definiteness we fix
$2m_{\tilde\chi^0_1}\sim (4/3)\ m_{\tilde \tau_1}
  \sim m_{\tilde\chi^+_1} (< m_{\tilde \mu_R})$
throughout the analysis, and vary the chargino mass.
The rest of the superpartners have fixed large masses
corresponding to the mSUGRA point $M_0=180$ GeV, 
$M_{1/2}=180$ GeV, $A_0=0$ GeV, $\tan\beta=44$ and $\mu>0$,
but we are not constrained to mSUGRA models only.
Our analysis will apply equally to gauge-mediated models with
a long-lived neutralino NLSP, as long as the relevant gaugino and slepton
mass relations are similar. Note that our choice of heavy
first two generation sleptons is very conservative.
A more judicious choice of their masses, namely
$m_{\tilde \mu_R}<m_{\tilde\chi^+_1}$, would lead to a
larger fraction of trilepton events, and as a result, a higher reach.
Furthermore, the gauginos would then decay via two-body modes
to first generation sleptons, and the resulting lepton spectrum
would be much harder, leading to a higher lepton efficiency.
Notice also that the $\tilde\chi^\pm_1\tilde\chi^0_2$
production cross-section is sensitive to the squark masses,
but since this is the only production process we are considering,
our results can be trivially rescaled to account for a different
choice of squark masses, or to include other production processes
as well.

Since the experimental signatures in our analysis contain only
soft leptons and tau jets, an important issue is whether one can
develop efficient combinations of Level 1 and Level 2 triggers to
accumulate these data sets without squandering all of the available
bandwidth.
We will not attempt to address this issue here; instead we will assume
100\% trigger efficiency for those signal events {\it which pass all
of our analysis and acceptance cuts}. We have nevertheless studied
the following set of triggers: 1) $\met>40$ GeV; 2) $p_T(l)>20$ GeV
and 3) $p_T(l)>10$ GeV, $p_T({\rm jet})>15$ GeV and $\met>15$ GeV;
with pseudorapidity cuts
$|\eta(e)|<2.0$, $|\eta(\mu)|<1.5$ and $|\eta(jet)|<4.0$.
We found that they are efficient in picking out about 90 \% of
the signal events in the channels with at least one lepton (see below).
Dedicated low $p_T$ tau triggers for Run II,
which may be suitable for the new tau
jet channels, are now being considered by both CDF \cite{CDF tau trigger}
and D0 \cite{D0 tau trigger}.

We used PYTHIA v6.115 and TAUOLA v2.5 for event generation.
We used the SHW v2.2 package \cite{SHW}, which
simulates an average of the CDF and D0 Run II detector performance.
In SHW tau objects are defined as jets with $|\eta|<1.5$, net
charge $\pm 1$, one or three tracks in a $10^\circ$
cone with no additional tracks in a $30^\circ$
cone, $E_T>5$ GeV, $p_T>5$ GeV, plus an electron rejection cut.
SHW electrons are required to have $|\eta|<1.5$, $E_T>5$ GeV,
hadronic to electromagnetic energy
deposit ratio $R_{h/e}<0.125$, and satisfy standard isolation cuts.
Muon objects are required to have $|\eta|<1.5$, $E_T>3$ GeV
and are reconstructed using Run I efficiencies. We use standard isolation
cuts for muons as well. Jets are required to have $|\eta|<4$, 
$E_T>15$ GeV. In addition we have added jet energy correction
for muons and the rather loose id requirement $R_{h/e}>0.1$.
We have also modified the TAUOLA program in order to correctly
account for the chirality of tau leptons coming from SUSY decays.

The reconstruction algorithms in SHW already include some
basic cuts, so we can define a reconstruction efficiency
$\epsilon_{rec}$ for the various types of objects: electrons,
muons, tau jets etc. We find that as we vary the chargino mass
from 100 to 140 GeV the lepton and tau jet reconstruction
efficiencies for the signal
range from 42 to 49 \%, and from 29 to 36\%, correspondingly.
The lepton efficiency may seem surprisingly low, but this is because
a lot of the leptons are very soft and fail the $E_T$ cut.
The tau efficiency is in good agreement with the results from 
Ref.~\cite{Hohlmann} and \cite{Leslie}, once we account
for the different environment, as well as cuts used in those
analyses.

The most important background issue in the new tau channels is the fake
tau rate. Several experimental analyses try to estimate it using
Run I data. Here we simulate the corresponding backgrounds to our
signal and use SHW to obtain the fake rate, thus
avoiding trigger bias \cite{Hohlmann}. We find that the tau fake rate
in $W$ production is 1.5\%, independent of the tau $p_T$,
which is in agreement with the findings of
Refs.~\cite{Hohlmann,Leslie,Eric}.

In the following we list our cuts for each channel.

In order to maximize the reach in the $lll\met$
channel, we apply the soft lepton $p_T$ cuts advertised in
Refs.~\cite{Barger}. We require a central lepton with 
$p_T>11$ GeV and $|\eta|<1.0$, and in addition two more leptons with
$p_T>7$ GeV and $p_T>5$ GeV.
Leptons have to be isolated: $I(l)<2$ GeV, where $I$ is
the total transverse energy contained in a cone of size
$\delta R=\sqrt{\Delta\varphi^2+\Delta\eta^2}=0.4$
around the lepton. 
We impose a dilepton invariant mass cut for same flavor,
opposite sign leptons: $|m_{ll}-M_Z|>10$ GeV and $|m_{ll}|>11$.
Finally, we impose an optional veto on additional jets
and require $\not\!\!\!E_T$ to be either
more than 20 GeV, or 25 GeV. This gives us a total of
four combinations of the $\not\!\!\!E_T$ cut and the jet veto (JV)
(A: $\met>20$ GeV, no JV;
 B: $\met>25$ GeV, no JV;
 C: $\met>20$ GeV, with JV;
 D: $\met>25$ GeV, with JV),
which we apply for all tau jet signatures later as well.

For our $ll\tau_h\not\!\!E_T$ analysis we impose cuts similar to the
stop search analysis in the $l^+l^-j\!\met$ channel \cite{CDFanalysis}:
two isolated ($I(l)<2$ GeV) leptons with $p_T>8$ GeV and $p_T>5$ GeV, and
one identified tau jet with $p_T(\tau_h)>15$ GeV. Again, we impose the above
invariant mass cuts for any same flavor, opposite sign dilepton pair.
This channel was also studied in Ref.~\cite{BCDPT-PRD}
with somewhat harder cuts on the leptons.
A separate, very interesting signature ($l^+ l^+ \tau_h\met$)
arises if the two leptons
have the same sign, since the background is greatly suppressed.
In fact, we expect this background to be significantly smaller than the
trilepton background! Roughly one third of the signal events in
the general $ll\tau_h$ sample are expected to have like-sign leptons.

For our $l\tau_h\tau_h\not\!\!E_T$ analysis we use some basic
identification cuts: two tau jets with $p_T>15$ GeV and  $p_T>10$ GeV
and one isolated lepton with $p_T>7$ GeV.

Finally, for the $\tau_h\tau_h\tau_h\not\!\!E_T$ signature
we only require three tau jets with
$p_T>15,10$ and 8 GeV, respectively.
%
%

\begin{figure}[t!]
\epsfysize=5.5in
\epsffile[0 180 350 670]{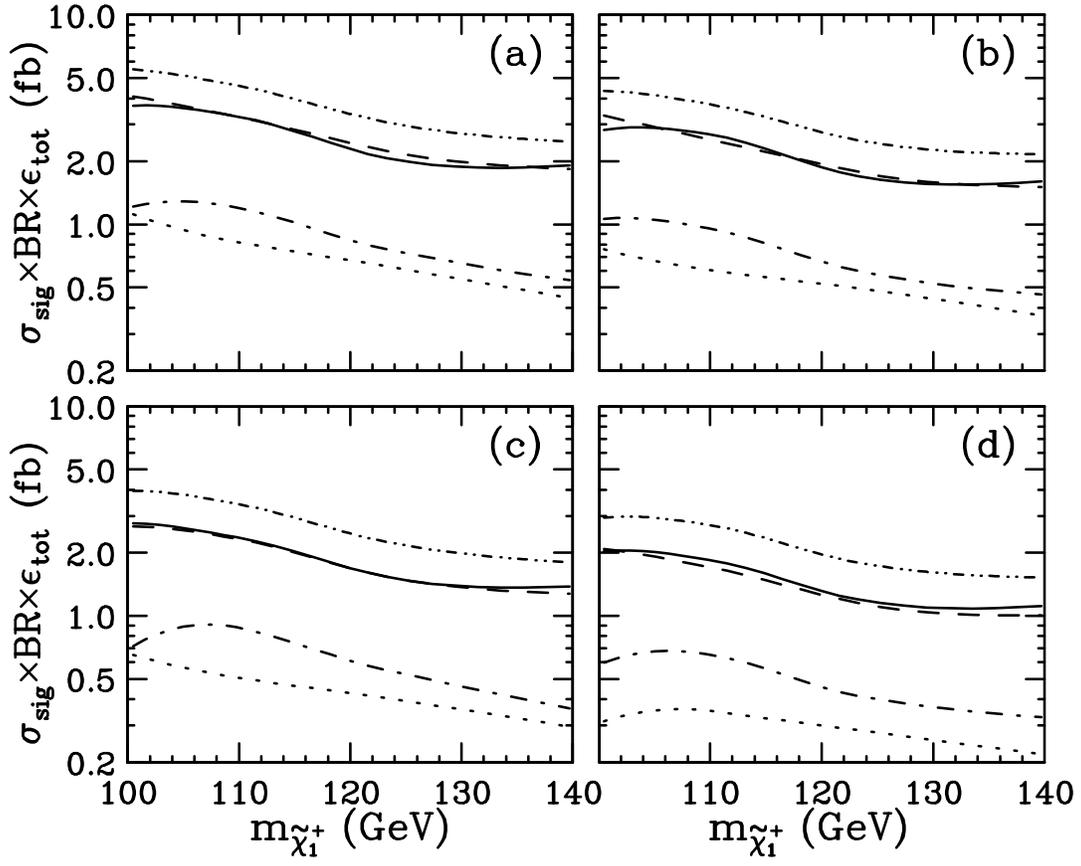}
\begin{center}
\parbox{5.5in}{
\caption[]{ Signal cross-section times branching ratio after cuts
for the five channels discussed in the text:
$lll\met$ (dotted),
$ll\tau_h\met$ (dashed),
$l^+ l^+ \tau_h\met$ (dot dashed),
$l\tau_h\tau_h\met$ (dot dot dashed) and
$\tau_h\tau_h\tau_h\met$ (solid);
and for various sets of cuts: (a) cuts A,
(b) cuts B, (c) cuts C and (d) cuts D.
\label{sigeff}}}
\end{center}
\end{figure}

One can get a good idea of the relative importance of the
different channels by looking at the corresponding signal samples
after the analysis cuts have been applied.
In Fig.~\ref{sigeff} we show the signal cross-sections
times the corresponding branching ratios times the total efficiency
$\epsilon_{tot}\equiv \epsilon_{rec}\epsilon_{cuts}$,
which accounts for both the detector acceptance $\epsilon_{rec}$
and the efficiency of the cuts $\epsilon_{cuts}$ (for each signal
point we generated $10^5$ events).
We see that the lines are roughly ordered according to the
branching ratios of three taus into the corresponding
final state signatures.
This can be understood as follows. The acceptance (which includes the
basic ID cuts in SHW) is higher for leptons than for $\tau$ jets.
Therefore, replacing a lepton with a tau jet in the
experimental signature costs us a factor of $\sim 1.5$ in acceptance,
due to the poorer reconstruction of tau jets, compared to leptons.
Later, however, the cuts tend to reduce the leptonic signal more than the
tau jet signal, mostly because the leptons are softer than the tau jets.
It turns out that these two effects mostly cancel each other, and the total
efficiency $\epsilon_{tot}$ is roughly the same for all channels.
Therefore the relative importance of each channel
will only depend on the tau branching ratios and the backgrounds. 
For example, in going from $lll$ to $ll\tau_h$,
one wins a factor of 5.5 from the branching ratio. 
Therefore the background to $ll\tau_h\met$
must be at least $5.5^2\sim 30$ times larger in order
for the clean trilepton channel to be still preferred.

We next turn to the discussion of the backgrounds involved.
We have simulated the following physics background processes:
$ZZ$, $WZ$, $WW$, $t\bar{t}$, $Z+{\rm jets}$, and $W+{\rm jets}$,
generating $4\times(10^6)$ and $2\times(10^7)$ events, respectively.
We list the results in Table~\ref{BKND}, where we show the total
background cross-section $\sigma_{BG}$ for each case A-D, as well as the
contributions from the individual processes, for case A.
All errors are purely statistical.
Our simulated background in the trilepton channel is higher
than previously found in Refs.~\cite{BCDPT-PRD,Barger}, which employed
ISAJET for event generation. Current versions of ISAJET
simulate the $WZ$ and $ZZ$ processes in the limit of zero $Z$ width.
We find that most of the $WZ$ and $ZZ$ background events
from PYTHIA contain an
off-shell $Z$ instead and pass the dilepton invariant mass cuts.
Unfortunately, neither ISAJET, nor PYTHIA contain the
$W\gamma$ interference contribution to $WZ$, so our
result still somewhat underestimates the $WZ$ trilepton background.
As we move to the channels with tau jets, the number
of events with {\em real} tau jets decreases, mostly
because of the smaller branching ratios of $W$ and $Z$ to taus.
However, the contribution from events with
fake tau jets increases, and for the $3\tau_h$
channel events with fakes are the dominant part of the
background. Our results for the $ll\tau_h$ and $l\tau_h\tau_h$
channels differ from those of Ref.~\cite{BT}, which assumed
a $p_T$-dependent fake tau rate of only 0.1-0.5\%.

We find that although the jet veto successfully removes the
$t\bar{t}$ background to the first three channels,
it also reduces the signal (see Fig.~\ref{sigeff}), and
hence does not improve signal-to-noise.
A higher $\not\!\!\!E_T$ cut also never seemed to help.
Indeed, the major backgrounds contain leptonic $t$ and $W$
decays and tend to have a lot of missing energy.
Notice that we did not account for fake leptons in
our $Z+{\rm jets}$ backgrounds to $3l$ and $l^+l^+\tau_h$,
since SHW does not provide a realistic simulation of those. 
The best way to estimate this background component
will be from Run II data (an analysis based on Run I data \cite{KM DP}
reveals that the $3l$ contribution from fakes is comparable to
our result in Table~\ref{BKND}, which only includes
real isolated leptons from heavy flavor jets).
We have a bit underestimated the total background
to the $3\tau_h$ channel by considering only processes with at least
one real tau in the event. We expect sizable contributions from
pure QCD multijet events, or $Wj\rightarrow jjj$, where
{\em all three} tau jets are fake. 
%
%

\begin{figure}[t!]
\epsfysize=5.5in
\epsffile[0 180 350 670]{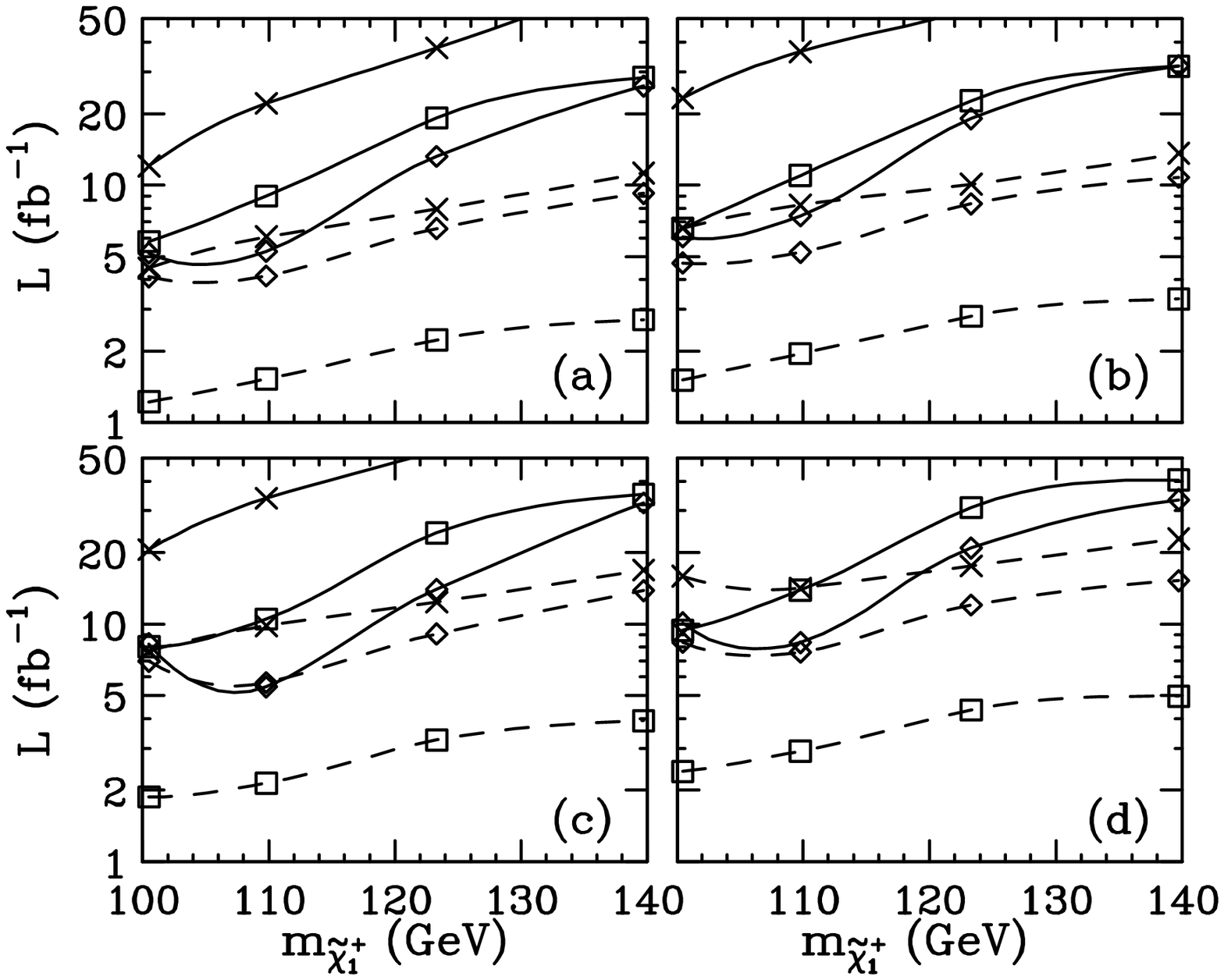}
\begin{center}
\parbox{5.5in}{
\caption[]{ The total integrated luminosity $L$ needed
for a $3\sigma$ exclusion (solid lines) or observation of
5 signal events (dashed lines), as a function of the chargino
mass $m_{\tilde\chi^+_1}$, for the three channels:
$lll\met$ ($\times$), $ll\tau_h\met$ ($\Box$) and
$l^+l^+\tau_h\met$ ($\Diamond$); and 
for various sets of cuts: (a) cuts A; (b) cuts B;
(c) cuts C and (d) cuts D.
\label{reach}}}
\end{center}
\end{figure}

A $3\sigma$ exclusion limit requires a total integrated luminosity
$ L(3\sigma)\ =\ 9\sigma_{BG}
\left\{\sigma_{sig}\ 
BR(\tilde\chi^+_1\tilde\chi^0_2\rightarrow X)
\ \epsilon_{tot}\right\}^{-2}.$
Notice that $L(3\sigma)$ depends linearly on the
background $\sigma_{BG}$ after cuts, but {\em quadratically} on
the signal branching ratios. This allows the jetty channels
to compete very successfully with the clean trilepton signature,
whose branching ratio is quite small.
In Fig.~\ref{reach} we show the Tevatron reach in the three best
channels: trileptons ($\times$), dileptons plus a tau jet ($\Box$)
and like-sign dileptons plus a tau jet ($\Diamond$).
We see that the two channels with tau jets have a much better
sensitivity compared to the usual trilepton signature.
Assuming that efficient triggers can be implemented,
the Tevatron reach will start exceeding LEP II
limits as soon as Run II is completed and the two collaborations
have collected a total of $4\ {\rm fb}^{-1}$ of data.
Considering the intrinsic difficulty of the SUSY scenario
we are contemplating, the mass reach for Run III is quite
impressive. One should also keep in mind that
we did not attempt to optimize our cuts for the new channels.
For example, one could use angular correlation cuts to suppress Drell-Yan,
transverse $W$ mass cut to suppress $WZ$ \cite{BTP-3L},
or (chargino) mass--dependent $p_T$ cuts for the leptons and tau jets,
to squeeze out some extra reach. In addition, the $ll\tau_h$ channel
can be explored at smaller values of $\tan\beta$ as well
\cite{BCDPT-PRD,Barger,KM DP}, since the two-body
chargino decays are preferentially to tau sleptons.
In that case, the clean trilepton channel still offers the best reach,
and a signal can be observed already in Run II. Then,
the tau channels will not only provide an important confirmation,
but also hint towards some probable values of the SUSY model parameters.

{\bf Acknowledgements.}
We would like to thank V.~Barger, J.~Conway, R.~Demina, L.~Groer,
J.~Nachtman, D.~Pierce, A.~Savoy-Navarro, M.~Schmitt and
A.~Turcott for useful discussions.
Fermilab is operated under DOE contract DE-AC02-76CH03000.

\newpage
\onecolumn

\begin{table*}[ht]
\caption{ Total background cross-section after cuts $\sigma_{BG}$
(in fb) for the various channels in cases A-D,
as well as the breakdown of the individual contributions for case A.
\label{BKND}}
\vspace{0.5in}
\renewcommand{\arraystretch}{1.5}
\begin{tabular}{cccccc}
$\sigma_{BG}$ (fb)
  & $lll\!\!\met$
    & $ll\tau_h\!\!\met$ 
      & $l^+l^+\tau_h\!\!\met$
        & $l\tau_h\tau_h\!\!\met$
          & $\tau_h\tau_h\tau_h\!\!\met$  \\ \hline\hline
$ZZ$           & 0.196 $\pm$ 0.028   
                    & 0.334 $\pm$ 0.036
                         & 0.094 $\pm$ 0.019
                              & 0.181 $\pm$ 0.027
                                   & 0.098 $\pm$ 0.020  \\ \hline
$WZ$           & 1.058 $\pm$ 0.052   
                    & 1.087 $\pm$ 0.053
                         & 0.447 $\pm$ 0.034
                              & 1.006 $\pm$ 0.051
                                   & 0.248 $\pm$ 0.025  \\ \hline
$WW$           & ---   
                    & 0.416 $\pm$ 0.061
                         & ---
                              & 0.681 $\pm$ 0.078
                                   & 0.177 $\pm$ 0.039  \\ \hline
$t\bar{t}$     & 0.300 $\pm$ 0.057   
                    & 1.543 $\pm$ 0.128
                         & 0.139 $\pm$ 0.038
                              & 1.039 $\pm$ 0.105
                                   & 0.161 $\pm$ 0.041  \\ \hline
$Z+{\rm jets}$ & 0.112 $\pm$ 0.079   
                    & 7.3 $\pm$ 0.6
                         & 0.168 $\pm$ 0.097
                              & 20.3 $\pm$ 1.1
                                   & 17.9 $\pm$ 1.0  \\ \hline
$W+{\rm jets}$ & ---    
                    & ---    
                         & ---    
                              & 37.2 $\pm$ 2.9
                                   & 6.1 $\pm$ 1.2  \\ \hline\hline
case A (total) & 1.67 $\pm$ 0.11   
                    & 10.7  $\pm$ 0.7
                         & 0.85 $\pm$ 0.11
                              & 60.4 $\pm$ 3.1
                                   & 24.7 $\pm$ 1.6  \\ \hline\hline
case B (total) & 1.49  $\pm$ 0.10
                    & 8.0  $\pm$ 0.5
                         & 0.76 $\pm$ 0.09
                              & 48.4 $\pm$ 2.8
                                   & 17.9 $\pm$ 1.4  \\ \hline\hline
case C (total) & 0.97  $\pm$ 0.07
                    & 6.4  $\pm$ 0.5
                         & 0.47 $\pm$ 0.06
                              & 39.6 $\pm$ 2.5
                                   & 13.8  $\pm$ 1.2  \\ \hline\hline
case D (total) & 0.83  $\pm$ 0.05
                    & 4.5  $\pm$ 0.3
                         & 0.40 $\pm$ 0.03
                              & 31.3 $\pm$ 2.4
                                   & 9.8  $\pm$ 1.0  \\ 
\end{tabular}
\end{table*}

\newpage

\newpage

\begin{references}

\bibitem{Tevatron searches}
See, e.g. M.~Carena {\em et al.},
preprint ANL-HEP-PR-97-98.

\bibitem{W to tau}
F.~Abe {\em et al.},
\prl 68 3398 1992 ; 
A.~Kotwal, ICHEP'98 Proceedings, Vancouver, Canada,
July 23-29, 1998; 
S.~Protopopescu, preprint FERMILAB-CONF-98-376-E.

\bibitem{top to tau}
M.~Hohlmann, preprint FERMILAB-CONF-96-330-E;
F.~Abe {\em et al.}, \prl 79 3585 1997 .

\bibitem{H+ to tau}
F.~Abe {\em et al.},
\prd 54 735 1996 ;
\prl 79 357 1997 .

\bibitem{Leptoquark to tau}
F.~Abe {\em et al.},
\prl 78 2906 1997 ; 
preprint FERMILAB-PUB-98-352-E.

\bibitem{BCDPT-PRL}
H.~Baer {\em et al.},
\prl 79 986 1997 . 

\bibitem{BCDPT-PRD}
H.~Baer {\em et al.},
\prd 58 075008 1998 . 

\bibitem{JW}
J.~Wells, \mpl 13 1923 1998 . 

\bibitem{DN}
B.~Dutta {\em et al.},
preprint OSU-HEP-98-4;
D.~Muller and S.~Nandi,
preprint OSU-HEP-98-8.

\bibitem{BT}
H.~Baer {\em et al.},
preprint FSU-HEP-990305.

\bibitem{Barger}
V.~Barger {\em et al.},
\plb 433 328 1998 ; 
V.~Barger and C.~P.~Kao,
preprint FERMILAB-PUB-98-342-T.

\bibitem{Run I 3L}
F.~Abe {\em et al.}, preprint FERMILAB-PUB-98/084-E;
B.~Abbott {\em et al.},
\prl 80 1591 1998 . 

\bibitem{CDF tau trigger}
Y. Seiya and A. Navoy-Savarro, talks given at the SUSY/Higgs Workshop,
Fermilab, 1998.

\bibitem{D0 tau trigger}
A. Turcott, private communication.

\bibitem{SHW}
J.~Conway, talk given at the SUSY/Higgs Workshop meeting,
Fermilab, May 14-16, 1998, additional information available at
{\tt www.physics.rutgers.edu/$\tilde{}$jconway/soft/shw/shw.html}.

\bibitem{Hohlmann}
M.~Hohlmann, Univ. of Chicago Ph.D. Thesis, 1997.

\bibitem{Leslie}
L.~Groer, Rutgers University Ph.D. Thesis, 1998. 

\bibitem{Eric}
E.~Smith, talk given at the Higgs and Supersymmetry conference,
Gainesville, FL, March 7-11, 1999.

\bibitem{CDFanalysis}
R.~Demina, talk given at the SUSY/Higgs Workshop meeting,
Fermilab, November 19-21, 1998.

\bibitem{KM DP}
K.~Matchev and D.~Pierce, preprint FERMILAB-Pub-99/078-T.

\bibitem{BTP-3L}
H.~Baer, M.~Drees, F.~Paige, P.~Quintana and X.~Tata,
preprint FSU-HEP-990509.

\end{references}
\end{document}